\documentclass[onecolumn]{IEEEtran}

\usepackage[numbers,sort&compress]{natbib}
\usepackage{amsthm}
\usepackage{amsmath}
\usepackage{amsfonts}
\usepackage[linesnumbered,ruled]{algorithm2e}

\newtheorem{definition}{Definition}
\newtheorem{lemma}{Lemma}

\newtheorem{theorem}{Theorem}

\begin{document}
\title{On the Local Correctabilities of Projective Reed-Muller Codes}
\author{\IEEEauthorblockN{Sian-Jheng Lin}\\
\IEEEauthorblockA{CAS Key Laboratory of Electro-magnetic Space Information\\
The School of Information Science and Technology\\
University of Science and Technology of China~(USTC)\\
Hefei, 230026, China\\
Email: sjlin@ustc.edu.cn}
}

\maketitle
\begin{abstract}
In this paper, we show that the projective Reed-Muller~(PRM) codes form a family of locally correctable codes~(LCC) in the regime of low query complexities. A PRM code is specified by the alphabet size $q$, the number of variables $m$, and the degree $d$. When $d\leq q-1$, we present a perfectly smooth local decoder to recover a symbol by accessing $\gamma\leq q$ symbols to the coordinates fall on a line. There are three major parameters considered in LCCs, namely the query complexity, the message length and the code length. This paper shows that PRM codes are shorter than generalized Reed-Muller~(GRM) codes in LCCs. Precisely, given a GRM code over a field of size $q$, there exists a class of shorter codes over a field of size $q-1$, while maintaining the same values on the query complexities and the message lengths.
\end{abstract}

\section{Introduction}
Projective Reed-Muller~(PRM) codes are first introduced by Lachaud~\cite{Lachaud:1988} in 1988, and the dimensions and minimum distances of PRM codes are determined by S\o rensen in \cite{104317}. From then on the properties of the codes have been studied extensively~\cite{Berger2001,Ding2002,995540,Ballet2014,7572207}. Several decoding approaches have been proposed \cite{6979865,NAKASHIMA2016}. In this paper, the local correctabilities of PRM codes are investigated. 

Locally decodable codes~(LDC)~\cite{TCS-030} are a class of error correcting codes allowing that each codeword symbol in the message part can be corrected probabilistically by accessing a few number of symbols in a corrected codeword with a randomized algorithm. If the local decoding property is not only available for the message part but also for the parity part, such codes are called locally correctable codes~(LCC). LDCs and LCCs have some important applications in complexity theory, cryptotic protocols and storage applications. 

To evaluate a class of codes for LDCs or LCCs, three major parameters will be considered, namely, the query complexity $\gamma$, the message length $k$, and the codeword length $n$. The query complexity indicates the number of codeword symbols that need to be accessed to recover a symbol of the codeword. The message length indicates the number of message symbols to be coded, and the codeword length is the number of symbols of the codeword. One of the major objectives in this research is to find out the shortest codes when the query complexities and the message lengths are specified. To date, a number of LDCs are proposed to serve for the different ratios between the query complexities and the message lengths. A typical family of LCCs are generalized Reed-Muller~(GRM) codes~\cite{1054127}, which are obtained by constructing the binary Reed-Muller codes~\cite{6499441,1057465} over a finite field. For low query complexities, Matching Vector~(MV) codes~\cite{Yekhanin:2008,Dvir:2010} are superior than GRM codes in LDCs, but MV codes are not belong to LCCs. For the case of large query complexities, GRM codes are not in this regime as their coding rates cannot higher than $1/2$. To overcome this issue, a number of codes are proposed in recent years~\cite{Guo:2013,Kopparty:2014,Kopparty:2016}. In this paper, the LCCs in low query complexities are considered. In \cite[Sec. 8.3]{TCS-030}, the author raise an open question regarding whether the existence of codes that are shorter than GRM codes in low query complexities. In this work, we answer this question by showing that PRM codes are shorter than GRM codes.

In this paper, a local decoder for PRM codes is proposed for low query complexities. This regime is previously occupied by GRM codes, and hence the results are compared with them. As tabulated in Table~\ref{table:a}, both codes are specified by three parameters $(q,d,m)$, where $q$ indicates the size of fields, $d$ indicates the degree of polynomials, and $m$ indicates the number of variables. From the table, given any GRM code with parameters $(q,d,m)$ and $d\leq q-2$, we can construct a PRM code with $(q-1,d,m)$~(Assume $q-1$ is also a prime power). It can be seen that in both codes, the query complexities and the message lengths are identical. However, the code length of GRM codes is $q^m$, that is always higher than $((q-1)^{m+1}-1)/(q-2)<(1+(q-2)^{-1})(q-1)^m=\Theta ((q-1)^m)$, the code length of PRM codes. This shows that the improvement is significant for small $q$. When $q=2$ (and $d=1$), PRM codes are Hadamard codes and the proposed algorithm is actually the same with the known local decoder for Hadamard codes. Table~\ref{table:a} also shows that PRM codes over a field of size $q-1$ have worse code distances than GRM codes over a field of size $q$. This can be seen as a tradeoff, whereas PRM codes provide higher coding rates in this case.

The proposed PRM local decoder is similar to the GRM local decoder. Given a coordinate $\mathbf{w}$ and a codeword, the decoder try to recover the symbol at coordinate $\mathbf{w}$ by accessing a few number of other symbols. Roughly speaking,  The decoder first randomly picks a coordinate $\mathbf{v}$, $\mathbf{v}\neq \mathbf{w}$, and queries the codeword symbols fall on the line passing through $\mathbf{w}$ and $\mathbf{v}$. After obtaining the responses for those symbols, the algorithm solves a single-variate polynomial $H(X)$ such that the evaluations of $H(X)$ meet the obtained values as many as possible, and then outputs the evaluation $H(0)$. The major differences between PRM codes and GRM codes are as follows. First, PRM local decoders consider the lines in projective space, and GRM local decoders consider the lines in affine space. Second, PRM local decoders can query the element at coordinate $\mathbf{v}$~(see \eqref{eq:L2}, the element $L_q$), and the symbols on the line form a codeword of extended Reed-Solomon codes. In contrast, GRM local decoders cannot query the element at $\mathbf{v}$, and those symbols form a codeword of Reed-Solomon codes.

The rest of this paper is organized as follows. Section \ref{sec:pre} introduces the definitions of LCCs and a number of traditional error correcting codes. Section~\ref{sec:LCPRM} introduces the proposed local decoder for PRM codes. Then the local correctabilities of PRM codes are shown in Section~\ref{sec:lcprm}. Section~\ref{sec:con} concludes this paper.

\begin{table}[!t]
\caption{\label{table:a}Parameters of Reed-Muller codes and projective Reed-Muller codes}
\label{table_example}
\centering
\begin{tabular}{|c||c|c|c|c|c|}
\hline
Codes & Restriction & Query complexity & Message length & Code length & Minimum distance\\
\hline
GRM & $d\leq q-2$ & $d+1$ & $\binom{m+d}{d}$ & $q^m$ & $(q-d)q^{m-1}$\\
\hline
PRM & $d\leq q-1$ & $d+1$& $\binom{m+d}{d}$ & $(q^{m+1}-1)/(q-1)$ & $(q-d)q^{m-1}$\\
\hline
\end{tabular}
\end{table}

\section{Preliminaries}\label{sec:pre}
\subsection{Nomenclature}
Let $\mathbb{F}_q$ denote a finite field of $q$ elements, for $q$ a prime power, and let $\mathbb{F}_q^*=\mathbb{F}_q\setminus\{0\}$. Let $\mathbf{X}=(X_1,\dots X_m)$ or $\mathbf{X}=(X_0,\dots X_m)$. An $m$-dimensional affine space over $\mathbb{F}_q$ is defined as
\[
\mathbb{A}^m(\mathbb{F}_q):=\{(a_1,\dots ,a_m)|a_j\in \mathbb{F}_q, j\in [m]\}.
\]
Further, an $m$-dimensional projective space is defined as
\[
\mathbb{P}^m(\mathbb{F}_q):=(\mathbb{A}^m(\mathbb{F}_q)\setminus \{\mathbf{0}\})/\sim,
\]
where $\sim$ is the equivalence relation defined as follows. Given $(a_1,\dots ,a_m)$ and $(b_1,\dots ,b_m)$, if there exists $\lambda\in \mathbb{F}_q^*$ such that $(a_1,\dots ,a_m) = (\lambda b_1,\dots ,\lambda b_m)$, then it can be written by
\[
(a_1,\dots ,a_m) \sim (b_1,\dots ,b_m),
\]
For simplicity, $\mathbb{A}^m$ and $\mathbb{P}^m$ are used to denote $\mathbb{A}^m(\mathbb{F}_q)$ and $\mathbb{P}^m(\mathbb{F}_q)$, respectively.

Let 
\[
\mathcal{H}_d^{m}=\mathbb{F}_q[X_1,\dots , X_m]\cup \{0\},
\]
where $\mathbb{F}_q[X_1,\dots , X_m]$ is a polynomial ring consisting of the homogeneous polynomials of degree $d$. For any $F(\mathbf{X})\in \mathcal{H}_d^{m}$, it is known that
\begin{equation}\label{eq:con1}
F(\lambda\mathbf{X})=\lambda^d F(\mathbf{X})\qquad \forall\lambda\in\mathbb{F}_q^*.
\end{equation}
\eqref{eq:con1} shows that $F(P)$ depends on the choice of the representative of $P=[p_1:p_2:\dots :p_m]\in \mathbb{P}^m$. To avoid the indeterminacies, it is necessary to specify the representative of the elements in $\mathbb{P}^m$. For $P\in \mathbb{P}^m$, we define
\[
\mathcal{D}(P)=p_i,
\]
where $i$ is the smallest integer such that $p_i\neq 0$. Then the representative of $P$ is defined by
\[
\mathcal{N}(P):=(0,\dots ,0,1,p_{i+1}',\dots ,p_{m}'),
\]
where each $p_j'=p_j/\mathcal{D}(P)$, for $j\geq i+1$. In addition, let
\[
\mathcal{N}(\mathbb{P}^m):=\{\mathcal{N}(P)|P\in \mathbb{P}^m\}.
\]

For $j\in \mathbb{N}$, $[j]=\{1,2,\dots ,j\}$ denotes a set of positive integers. For $\mathbf{x}, \mathbf{y}\in \mathbb{F}_q^n$, $\Delta(\mathbf{x}, \mathbf{y})$ denotes the relative Hamming distance between $\mathbf{x}$ and $\mathbf{y}$. That is, the ratio of different elements between $\mathbf{x}$ and $\mathbf{y}$. For $\mathbf{x}\in \mathbb{F}_q^m$ with an integer $m$, $\mathbf{x}[i]$ denotes the $i$-th symbol of $\mathbf{x}$, and $\mathbf{x}| _{S}$ denotes $\mathbf{x}$ restricted to symbols indexed by $S\subset [j]$.

\subsection{Locally correctable codes}
A class of codes of message length $k$ and codeword length $n$ is called $(\gamma, \delta, \epsilon)$-locally correctable, if for each received codeword $\mathbf{y}$ with up to $\delta n$ erasures, each symbol $\mathbf{y}[i]$, $i\in [n]$, can be recovered with probability $1-\epsilon$, by accessing at most $\gamma$ symbols chosen by a randomized algorithm. A formal definition of locally correctable codes is addressed below.
\begin{definition}\label{lemma:LCC}
(Locally Correctable Code~(LCC)) A code $\mathcal{C}\subset \mathbb{F}_q^n$ is $(\gamma, \delta, \epsilon)$-locally correctable, if there exists a randomized algorithm $\mathcal{A}$ such that for each pair $(\mathbf{c}\in\mathcal{C}, \mathbf{y}\in \mathbb{F}_q^n)$ and $\Delta (\mathbf{c}, \mathbf{y})\leq \delta$, 
\[
\mathrm{Pr}[\mathcal{A}(\mathbf{y}, i) = \mathbf{c}(i)]\geq 1- \epsilon
\]
holds for each $i\in [n]$, and further $\mathcal{A}$ accesses at most $\gamma$ symbols of $\mathbf{y}$.
\end{definition}
In Definition~\ref{lemma:LCC}, if the local decoding property is available for $i\in [k]$, such codes are called locally decodable codes~(LDCs). Clearly, LCC is a subset of LDC. Since this paper only considers LCCs, the details about LDCs are omitted. 

A local decoder $\mathcal{A}$ consists of two parts, namely the randomized query algorithm $Q$ and the deterministic reconstruction algorithm $R$. A local decoder with query complexity $\gamma$ is called perfectly smooth, if the following requirements are satisfied~\cite{Trevisan04, Hemenway13}. First, the deterministic reconstruction algorithm can recover any codeword symbol by accessing other $\gamma$ symbols within the codeword at most. Second, the randomized query algorithm meets the requirement that, each query is uniformly distributed in the codeword. This means that for each symbol, all other symbols has equal chance of being selected in the set of queries. A formal definition is addressed below.

\begin{definition}\label{lemma:psmooth}
(Perfectly Smooth Decoder) For a code $\mathcal{C}\subset \mathbb{F}_q^n$ with query complexity $\gamma$, a local decoder $\mathcal{A}$ consists of a randomized query algorithm $Q$ and a deterministic reconstruction algorithm $R:\mathbb{F}_q^{\gamma} \times [n] \rightarrow \mathbb{F}_q$. For each $c \in \mathcal{C}$ and a point $i\in [n]$, $Q$ reads $i$ and generates a set of queries $Q(i)$ with $|Q(i)|\leq \gamma$. Next, $R$ reads $c|_{Q(i)}$ and $i$ to recover $c[i]$. The local decoder is perfectly smooth if the following conditions hold.
\begin{enumerate}
\item For each $c \in \mathcal{C}$ and $i \in [n]$, 
\[
R( c|_{Q(i)} , i) = c[i].
\]
\item For each $i\in [n]$, each query in $Q(i)$ is uniformly distributed over $[n]$.
\end{enumerate}
\end{definition}

\subsection{Error correcting codes}
A number of error correcting codes are introduced in this subsection. 

\subsubsection{Reed-Solomon codes}
$(n,k)$ Reed-Solomon~(RS) codes~\cite{1960} over $\mathbb{F}_q$ treat the message as a single-variate polynomial of degrees less than $k$, and the codeword is generated by evaluating the polynomial to $n= q$ fixed points. In addition, the extended Reed-Solomon code~(or called doubly-extended Reed-Solomon code) is constructed by appending an extra symbol to the codeword of $(q,k)$ RS codes, where the extra symbol is the coefficient of the polynomial at degree $k-1$. The formal definition is given as follows.
\begin{definition}\label{lemma:RS}
The Reed-Solomon code over $\mathbb{F}_q$ of order $d$ and length $n=q$ is defined by
\[
\mathbf{RS}_q (d) = \{(F(\lambda))_{\lambda\in \mathbb{F}_q}|F(X)\in \mathbb{F}_q [X],\deg F\leq d\}.
\]
The extended Reed-Solomon code is defined by
\[
\mathbf{ERS}_q (d) = \{(F(\lambda_0),\dots ,F(\lambda_{q-1}), F(\lambda_{\infty}))|F(X)\in \mathbb{F}_q [X],\deg F\leq d\},
\]
where $F(\lambda_{\infty})$ denotes the coefficient of $X^d$.
\end{definition}
RS codes are maximum distance separable~(MDS) codes, which possess the optimal trade-off between minimum distances and the size of redundancies. $(n,k)$ RS codes are able to correct up to $E$ errors and $S$ erasures, as long as $2E+S \leq n-k$. A number of typical decoders, such as Berlekamp-Welch algorithm~\cite{WelchBerlekamp1986} and Berlekamp-Massey algorithm~\cite{1054260}, can be used to decode RS codes. In addition, the efficient decoders for extended RS codes are known~\cite{535795,510064}. 

\subsubsection{Generalized Reed-Muller codes}\label{sec:GRM}
Generalized Reed-Muller~(GRM) codes~\cite{1054127} are a family of linear error-correcting codes by constructing binary Reed-Muller~(RM) codes~\cite{1057465} over a finite field. In GRM codes $\mathbf{GRM}_d (m,q)$, the message is determined by an $m$-variate polynomial of degree at most $d$ over $\mathbb{F}_q$, and the codeword is defined as the evaluations of the polynomial at $\mathbb{F}_q^m$. The formal definition is as follows.
\begin{definition}\label{lemma:GRM}
The generalized Reed-Muller code over $\mathbb{F}_q$ of order $d$ and length $n=q^m$ is defined by
\[
\mathbf{GRM}_q (d,m) = \{(F(A))_{A\in \mathbb{A}^m}|F(\mathbf{X})\in \mathbb{F}_q [X_1,\dots ,X_m],\deg F\leq d\},
\]
where each $A_i\in \mathbb{A}^m$.
\end{definition}
The code dimension and the minimum distance of GRM are known~\cite[p.72]{BLAKE19751}. In particular, $\mathbf{GRM}_q (d,1)$ is Reed-Solomon codes, and $\mathbf{GRM}_2 (1,m)$ is a punctured version of Hadamard codes.

When $d\leq q-2$, the GRM codes form a typical family of locally correctable codes with the query complexity $d+1$, the message length $k=\binom{m+d}{d}$ and the code length $q^m$. To recover a symbol at $\mathbf{w}\in \mathbb{A}^m$, the local decoder randomly picks a vector $\mathbf{v}\in \mathbb{A}^m\setminus \mathbf{w}$ in uniform distribution and queries $d+1$ symbols fall on 
\begin{equation}\label{eq:RMQuery}
L=\{\mathbf{w}+\lambda \mathbf{v}|\lambda\in \mathbb{F}_q^*\}.
\end{equation}
Those symbols form a codeword of $(q,d+1)$ RS codes. By applying RS decoding algorithm on it, one can obtain a single-variate polynomial $H(X)$ with degree at most $d$, and then the decoder returns $H(0)$.

\subsubsection{Projective Reed-Muller codes}
Projective Reed-Muller~(PRM) codes~\cite{Lachaud:1988} are a variant of GRM codes. In PRM codes $\mathbf{PRM}_q (d,m)$, the message is determined by an $(m+1)$-variate homogeneous polynomial of degree $d$ over $\mathbb{F}_q$, and the codeword is obtained by evaluating the polynomial in a $(m+1)$-dimensional projective space. The PRM code is defined as follows.
\begin{definition}\cite{104317}\label{lemma:PRM}
The projective Reed-Muller code over $\mathbb{F}_q$ of order $d$ and length $n=(q^{m+1}-1)/(q-1)$ is defined by
\[
\mathbf{PRM}_q (d,m) = \{(F(P))_{P\in \mathcal{N}(\mathbb{P}^{m+1})}|F(\mathbf{X})\in \mathcal{H}_d^{m+1}\}.
\]
\end{definition}
The code dimension and the minimum distance of GRM are known~\cite{104317}. Notably, $\mathbf{PRM}_q (d,1)$ is extended Reed-Solomon codes, and $\mathbf{PRM}_2 (1,m)$ is Hadamard codes. In this case $\mathbf{PRM}_2 (1,m)$, the proposed local decoder is the same with the local decoder for Hadamard codes. 

\section{Perfectly smooth decoder for PRM codes}\label{sec:LCPRM}
\begin{algorithm}[t]
\caption{\label{alg:Q} Randomized query algorithm for PRM codes}
\KwIn{$\mathbf{w}\in \mathbb{P}^{m+1}$ and $d$}
\KwOut{
$\Lambda=(\lambda_i\in \mathbb{F}_q^*\cup \{\infty\})_{i\in [d+1]}$, 
$L=(L_i\in \mathcal{N}(\mathbb{P}^{m+1}))_{i\in [d+1]}$ and $D=(D_i\in \mathbb{F}_q)_{i\in [d+1]}$}
Let $S$, $|S|=d+1$, be an arbitrary subset of $\mathbb{F}_q^*\cup \{\infty\}$, and $\Lambda$ is constructed by ordering the elements of $S$ in random permutation.\\
Choose $\mathbf{v}\in \mathbb{P}^{m+1}\setminus \{\mathbf{w}\}$ randomly in uniform distribution.\\
\For {$i=0,\dots ,q-1$}
{
$L_i=\left\{
\begin{matrix}
\mathbf{v} & \text{If }\lambda_i=\infty\\ 
\mathcal{N}(\mathbf{w}+\lambda_i \mathbf{v}) & \text{otherwise}
\end{matrix}\right.
$\\
$D_i=\left\{
\begin{matrix}
1 & \text{If }\lambda_i=\infty\\ 
\mathcal{D}(\mathbf{w}+\lambda_i \mathbf{v}) & \text{otherwise}
\end{matrix}\right.
$
}
\Return $\Lambda$, $L$ and $D$.
\end{algorithm}

\begin{algorithm}[t]
\caption{\label{alg:R} Deterministic reconstruction algorithm for PRM codes}
\KwIn{$\Lambda$, $(e_i=F(L_i))_{i\in [d+1]}$, $D$ and $\mathbf{w}$}
\KwOut{$F(\mathbf{w})$}
Find out a polynomial $H(X)$, $\deg H \leq d$, such that
\begin{equation}
\begin{aligned}
H(\lambda_i)&=\mathcal{D}_i^d\times e_i\qquad i\in [d+1],
\end{aligned}
\end{equation}
where $H(\infty)$ denotes the coefficient of $X^d$.\\
\Return $H(0)$.
\end{algorithm}

In this section, a $(d+1)$-query perfectly smooth decoder for $\mathbf{PRM}_q (d,m)$, $d\leq q-1$, is proposed. The approach is similar to the local decoder for GRM codes. By following Definition~\ref{lemma:psmooth}, the decoder is denoted as a pair of algorithms $(Q,A)$. Given a codeword $((F(P))_{P\in \mathcal{N}(\mathbb{P}^{m+1})})\in \mathbf{PRM}_q (d,m)$ and a point $\mathbf{w}\in \mathbb{P}^{m+1}$, the value $F(\mathbf{w})$ can be recovered via the following steps. First, the decoder randomly picks $\mathbf{v}\in \mathcal{N}(\mathbb{P}^{m+1})\setminus \mathbf{w}$ in uniform distribution. Then we consider a line passing through $\mathbf{w}$ and $\mathbf{v}$: 
\begin{equation}\label{eq:L2}
L_\mathbf{w}(\mathbf{v}):=\{\mathcal{N}(\mathbf{w}+\lambda\mathbf{v})|\lambda\in \mathbb{F}_q\}\cup \{\mathbf{v}\}.
\end{equation}
Notably, $L_\mathbf{w}(\mathbf{v})$ includes $\mathbf{v}$, which is not in the set~\eqref{eq:RMQuery} for RM codes. Let $S$ denote an arbitrary subset of $\mathbb{F}_q^*\cup \{\infty\}$, and $|S|=d+1$. The decoder queries $d+1$ symbols fall in 
\[
\{\mathcal{N}(\mathbf{w}+\lambda\mathbf{v})|\lambda\in S\}.
\]
Notably, we define $\mathcal{N}(\mathbf{w}+\lambda\mathbf{v})=\mathbf{v}$ if $\lambda=\infty$. The obtained values are denoted as 
\begin{equation}\label{eq:e_}
\{e_\lambda=F(\mathcal{N}(\mathbf{w}+\lambda\mathbf{v}) )|\lambda\in S\}.
\end{equation}
Second, the decoder solves a single-variate polynomial $H(X)$, $\deg H\leq d$ such that
\begin{equation}\label{eq:H}
H(\lambda)=D_\lambda^d\times  e_\lambda \qquad \lambda\in S,
\end{equation}
where 
\begin{equation}\label{eq:H2}
D_\lambda=\mathcal{D}(\mathbf{w}+\mathbf{v}\cdot \lambda)\in\mathbb{F}_q. 
\end{equation}
Further, if $\infty\in S$, \eqref{eq:H} has the equation
\begin{equation}\label{eq:le2}
H(\infty)=D_\infty^d\times  e_\infty,
\end{equation}
which means that the coefficient of $X^d$ is equivalent to $D_\infty^d\times  e_\infty$. After obtaining $H(X)$, the decoder returns $H(0)=F(\mathbf{w})$. The details are addressed in Algorithm~\ref{alg:Q} and Algorithm~\ref{alg:R}. The following shows that the decoder meets the requirements of the perfectly smooth decoder in Definition~\ref{lemma:psmooth}.

To verity the first requirement, we show that the set \eqref{eq:e_} can be considered as a set of evaluations of a single-variate polynomial. Thus, $F(\mathbf{w})$ can be recovered via the decoding algorithm of extended RS codes, and the first requirement holds. To simplify the derivations, another formulation of \eqref{eq:H} is given as follows. Based on the fact that $F$ is a homogeneous polynomial, we have
\begin{equation}\label{eq:ls}
\begin{aligned}
&F(\mathcal{N}(\mathbf{w}+\mathbf{v}\cdot \lambda))\\
=&D_\lambda^{-d}\times  F(D_\lambda\times\mathcal{N}(\mathbf{w}+\mathbf{v}\cdot \lambda))\\
=&D_\lambda^{-d}\times F(\mathbf{w}+\mathbf{v}\cdot \lambda).\\
\end{aligned}
\end{equation}
where $D_\lambda$ is as defined in~\eqref{eq:H2}. Thus, \eqref{eq:H} can be written as
\begin{equation}\label{eq:le2.5}
H(\lambda)=F(\mathbf{w}+\mathbf{v}\cdot \lambda) \qquad \lambda\in S.
\end{equation}
That is, it is equivalent to show that $\{F(\mathbf{w}+\mathbf{v}\cdot \lambda )|\lambda \in \mathbb{F}_q\}\cup \{F(\mathbf{v})\}$ forms a set of evaluations of a single variate polynomial. This statement is proved below.

\begin{lemma}\label{lemma:LC}
For any $\mathbf{v},\mathbf{w}\in \mathcal{N}(\mathbb{P}^{m+1})$, $\mathbf{v}\neq \mathbf{w}$, and any $F(\mathbf{X})\in \mathcal{H}_d^{m+1}$, $d\leq q-1$, there exists a single-variate polynomial $H(X)$, $\deg H\leq d$ such that
\begin{equation}\label{eq:le1}
H(\lambda)=F(\mathbf{w}+\mathbf{v}\cdot \lambda)\qquad \lambda\in \mathbb{F}_q,
\end{equation}
\begin{equation}\label{eq:le1.2}
H(\infty)=F(\mathbf{v}),
\end{equation}
where $H(\infty)$ denotes the coefficient of $H(X)$ at degree $d$.
\end{lemma}
\begin{proof}
The homogeneous polynomial $F(\mathbf{X})$ is written by
\[
F(\mathbf{X})=\sum_{d_0+\dots +d_m=d} \gamma_{d_0,\dots ,d_m}\prod_{j=0}^{m} X_j^{d_j}.
\]
By plugging $(\mathbf{w}X_0+\mathbf{v}X_1)$ into $F(\mathbf{X})$, we obtain
\begin{equation}\label{eq:L3s}
\begin{aligned}
F(\mathbf{w}X_0+\mathbf{v}X_1)&=F(w_0X_0+v_0X_1,\dots ,w_mX_0+v_mX_1)\\
&=\sum_{d_0+\dots +d_m=d} \gamma_{d_0,\dots ,d_m}\prod_{j=0}^{m} (w_jX_0+v_jX_1)^{d_j}\\
&=F_{\mathbf{w},\mathbf{v}}(X_0, X_1).
\end{aligned}
\end{equation}
From \eqref{eq:L3s}, it can be observed that $F_{\mathbf{w},\mathbf{v}}(X_0, X_1)$ is also a homogeneous polynomial of degree $d$ in $X_0$ and $X_1$. It can be observed that
\begin{equation}\label{eq:L32}
(F_{\mathbf{w},\mathbf{v}}(P))_{P\in\mathcal{N}(\mathbb{P}^2)}\in \mathbf{PRM}_q (d, 2),
\end{equation}
which is equivalently the extended RS codes. 

In \eqref{eq:L32}, the set of evaluation points can be written as
\begin{equation}\label{eq:L321}
\mathcal{N}(\mathbb{P}^2)=\{(1,\lambda)|\lambda\in \mathbb{F}_q\}\cup \{(0,1)\}.
\end{equation}
In the following, we show that
\begin{equation}\label{eq:L33}
H(X)=F_{\mathbf{w},\mathbf{v}}(1,X)=\sum_{d_0+\dots +d_m=d} \gamma_{d_0,\dots ,d_m}\prod_{j=0}^{m} (w_j+v_jX)^{d_j}
\end{equation}
meets \eqref{eq:le1} and \eqref{eq:le1.2}. It can be seen that $\deg H\leq d$. To verify \eqref{eq:le1.2}, $(X_0,X_1)=(0,1)$ is plugged into \eqref{eq:L3s} to obtain
\begin{equation}\label{eq:L33s}
\begin{aligned}
F(\mathbf{v})&=F(v_0,\dots ,v_m)=\sum_{d_0+\dots +d_m=d} \gamma_{d_0,\dots ,d_m}\prod_{j=0}^{m} v_j^{d_j}=F_{\mathbf{w},\mathbf{v}}(0,1),
\end{aligned}
\end{equation}
and one can verify that $F(\mathbf{v})$ is equivalent to the coefficient of $H(X)$ at degree $d$, and this shows that \eqref{eq:le1.2} holds. 

To verify \eqref{eq:le1}, $\lambda\in \mathbb{F}_q$ is plugged into $H(X)$, resulting in
\begin{equation}\label{eq:L34}
H(\lambda)=F_{\mathbf{w},\mathbf{v}}(1, \lambda)=F(\mathbf{w}+\mathbf{v}\cdot \lambda).
\end{equation}
This completes the proof.
\end{proof}

Next, the second requirement of the perfectly smooth decoder is considered. First of all, we show the following result.
\begin{lemma}\label{lemma:L1}
$L_\mathbf{w}(\mathbf{v})$ includes $q+1$ distinct elements of $\mathbb{P}^{m+1}$.
\end{lemma}
\begin{proof}
It is equivalent to show the following two statements:
\begin{equation}\label{eq:state2}
\mathbf{v}\notin\{\mathcal{N}(\mathbf{w}+\lambda \mathbf{v})|\lambda\in\mathbb{F}_q\},
\end{equation}
\begin{equation}\label{eq:state1}
\mathcal{N}(\mathbf{w}+\lambda_0 \mathbf{v})\neq\mathcal{N}(\mathbf{w}+\lambda_1 \mathbf{v})\qquad \forall\lambda_0,\lambda_1\in \mathbb{F}_q, \lambda_0\neq\lambda_1.
\end{equation}
Those can be proved by contradictions. To verify \eqref{eq:state2}, assume there exists $\lambda_0\in \mathbb{F}_q$ such that
\begin{equation}\label{eq:state2.1}
\mathbf{v}=\mathcal{N}(\mathbf{w}+\lambda_0 \mathbf{v}).
\end{equation}
\eqref{eq:state2.1} implies there exists $\gamma\in \mathbb{F}_q^*$ such that
\begin{equation}\label{eq:state2.2}
\gamma\mathbf{v}=\mathbf{w}+\lambda_0 \mathbf{v}
\Rightarrow \mathbf{w}=(\gamma-\lambda_0)\mathbf{v}.
\end{equation}
As $\mathbf{w}\neq \mathbf{0}$, we have $(\gamma-\lambda_0)\neq 0$, and \eqref{eq:state2.2} can be written as
\[
\mathcal{N}(\mathbf{w})=\mathcal{N}(\mathbf{v}),
\]
which contradicts the definition $\mathbf{w}\neq\mathbf{v}$. Thus, the assumption is false and \eqref{eq:state2} is proved. 

To verify \eqref{eq:state1}, assume there exists $\delta_0, \delta_1\in \mathbb{F}_q$, $\delta_0\neq \delta_1$, such that
\begin{equation}\label{eq:state1.1}
\mathcal{N}(\mathbf{w}+\delta_0 \mathbf{v})=\mathcal{N}(\mathbf{w}+\delta_1 \mathbf{v}).
\end{equation}
\eqref{eq:state1.1} implies that there exists $\gamma\in \mathbb{F}_q^*$ such that
\begin{equation}\label{eq:state1.2}
\begin{aligned}
\mathbf{w}+\delta_0 \mathbf{v}=\gamma(\mathbf{w}+\delta_1 \mathbf{v})\\
\Rightarrow (1-\gamma)\mathbf{w}=(\gamma\delta_1-\delta_0)\mathbf{v}.
\end{aligned}
\end{equation}
In \eqref{eq:state1.2}, as $\mathbf{w}, \mathbf{v}\neq \mathbf{0}$ and $\delta_0 \neq \delta_1$, the only possible result is $(1-\gamma)\neq 0$ and $(\gamma\delta_1-\delta_0)\neq 0$, and hence 
\[
\mathcal{N}(\mathbf{w})=\mathcal{N}(\mathbf{v}),
\]
which contradicts the definition $\mathbf{w}\neq\mathbf{v}$. Thus, the assumption is false. This completes the proof.
\end{proof}
With Lemma~\ref{lemma:L1}, the second requirement is shown as follows.
\begin{lemma}\label{lemma:L2}
For any $\mathbf{w}, \mathbf{p}\in \mathcal{N}(\mathbb{P}^{m+1})$ and $\mathbf{w}\neq\mathbf{p}$, $L_\mathbf{w}(\mathbf{v})$ is constructed by choosing $\mathbf{v}\in \mathcal{N}(\mathbb{P}^{m+1})\setminus \{\mathbf{w}\}$ uniformly, then 
\[
\mathrm{Pr}[\mathbf{p}\in L_\mathbf{w}(\mathbf{v})]=(q-1)/(q^m-1).
\]
\end{lemma}
\begin{proof}
From the definition \eqref{eq:L2}, when $\mathbf{p}\in L_\mathbf{w}(\mathbf{v})$, we have
\begin{equation}\label{eq:L2.1}
\mathbf{p}\in\{\mathbf{v}\},
\end{equation}
or
\begin{equation}\label{eq:L2.20}
\mathbf{p}\in\{\mathcal{N}(\mathbf{w}+\lambda \mathbf{v})|\lambda\in \mathbb{F}_q^*\}.
\end{equation}
\eqref{eq:L2.20} implies that there exists $\gamma, \lambda_0\in \mathbb{F}_q^*$ such that
\begin{equation}\label{eq:L2.2}
\begin{aligned}
&\gamma\mathbf{p}=\mathbf{w}+\lambda_0 \mathbf{v}\\
\Rightarrow &\mathbf{w}-\gamma\mathbf{p}= -\lambda_0\mathbf{v}\\
\Rightarrow &\mathcal{N}(\mathbf{w}-\gamma\mathbf{p})=\mathbf{v}.\\
\end{aligned}
\end{equation}
From \eqref{eq:L2.1} and \eqref{eq:L2.2}, we have
\begin{equation}\label{eq:L2.3}
\begin{aligned}
&\mathbf{v}\in \{\mathcal{N}(\mathbf{w}-\gamma\mathbf{p})|\gamma\in\mathbb{F}_q^*\}\cup \{\mathbf{p}\}\\
\Rightarrow&\mathbf{v}\in L_\mathbf{w}(\mathbf{p})\setminus\{\mathbf{w}\}.
\end{aligned}
\end{equation}
\eqref{eq:L2.3} shows that 
\begin{equation}\label{eq:L2.4}
\mathrm{Pr}[\mathbf{p}\in L_\mathbf{w}(\mathbf{v})]=\mathrm{Pr}[\mathbf{v}\in L_\mathbf{w}(\mathbf{p})\setminus\{\mathbf{w}\}].
\end{equation}
From Lemma~\ref{lemma:L1}, $|L_\mathbf{w}(\mathbf{p})\setminus\{\mathbf{w}\}|=q$. Since $\mathbf{v}$ is chosen uniformly in $\mathcal{N}(\mathbb{P}^{m+1})\setminus \{\mathbf{w}\}$, we have
\begin{equation}\label{eq:L2.5}
\mathrm{Pr}[\mathbf{v}\in L_\mathbf{w}(\mathbb{P})\setminus\{\mathbf{w}\}]=q/(n-1)=(q-1)/(q^m-1).
\end{equation}
From \eqref{eq:L2.4} and \eqref{eq:L2.5}, the proof is completed.
\end{proof}

Lemma~\ref{lemma:L2} shows that for each element of $\mathcal{N}(\mathbb{P}^{m+1})\setminus \{\mathbf{w}\}$ has equal probability to be chosen in $L_\mathbf{w}(\mathbf{v})$. In Algorithm~\ref{alg:Q}, the order of queries is randomly permuted, and hence the $j$-th query, $j\in [q]$, is uniformly distributed in $\mathcal{N}(\mathbb{P}^{m+1})\setminus \{\mathbf{w}\}$. Thus the proposed decoder meets the second requirement of perfectly smooth decoders.

It is valuable to notice that, the permutation of queries~(Step 1 of Algorithm~\ref{alg:Q}) is necessary. If the permutation of queries is omitted, the array of queries may be given by 
\begin{equation}\label{eq:aa}
(L_i=\mathcal{N}(\mathbf{w}+\omega_i\mathbf{v}))_{i\in [q]},
\end{equation}
where $\{\omega_i\}_{i\in [q]}$ denotes the $q$ elements of $\mathbb{F}_q$. We note that the list~\eqref{eq:aa} cannot satisfy the second requirement for the perfectly smooth decoder~(Definition~\ref{lemma:psmooth}), though this problem has not appeared in GRM codes. To see this, let $V(\mathbf{X})=\mathcal{N}(\mathbf{w}+\lambda\mathbf{X})$ with domain/codomain $\mathcal{N}(\mathbb{P}^{m+1})$. The notation $\bar{w}$ denotes the smallest integer such that $\mathbf{w}[\bar{w}]\neq 0$, and $\bar{v}$ denotes the smallest integer such that $\mathbf{v}[\bar{v}]\neq 0$. Then when $\bar{w}<\bar{v}$, there exists $\mathbf{v}'\neq\mathbf{v}$ such that $V(\mathbf{v})=V(\mathbf{v}')$. That is, $\mathbf{v}'=\mathbf{w}+(1+\lambda)\mathbf{v}$. Since $V$ is not bijective, the image of $V$ is a proper subset of $\mathcal{N}(\mathbb{P}^{m+1})$, and hence the query $L_i$ is not uniformly distributed in $\mathcal{N}(\mathbb{P}^{m+1})$.

\section{Local correctabilities of PRM codes}\label{sec:lcprm}
Based on the proposed local decoder, this section shows that PRM codes form a family of locally correctable codes. In this section, the received codewords are corrupted, as opposed to Section~\ref{sec:LCPRM} that the codewords do not have errors. The results in this section follow the prior results for RM codes~\cite[Sec. 2.2]{TCS-030}.

\begin{theorem}\label{themoem:LC1}
The projective Reed-Muller code~$\mathbf{PRM}_q (d,m)$, $d\leq q-1$, is $(d+1, \delta , (d+1) \delta )$-locally correctable for all $\delta$. 
\end{theorem}
\begin{proof}
The algorithm is the same with the decoder in Section~\ref{sec:LCPRM}, except that the corrupted codewords are considered. Given a codeword generated by a polynomial $F(\mathbf{X})$ with $\delta n$ errors and a point $\mathbf{w}\in \mathcal{N}(\mathbb{P}^{m+1})$, the objective is to recover the value $F(\mathbf{w})$ by accessing at most $d+1$ symbols of $\mathbf{y}$. First, the decoder calls Algorithm~\ref{alg:Q} to obtain the list of queries $L$. After obtaining the symbol values corresponding to $L$, the decoder calls Algorithm~\ref{alg:R} to obtain the result. Since each query is uniformly distributed, the probability that all queries are not corrupted is at least $1-(d+1)\delta$.
\end{proof}

\begin{theorem}\label{themoem:LC2}
The projective Reed-Muller code~$\mathbf{PRM}_q (d,m)$, $d\leq \sigma q-1$, is $(q, \delta , 2\delta /(1-\sigma))$-locally correctable for all $\delta$. 
\end{theorem}
\begin{proof}
The algorithm is a modification of the decoder in Section~\ref{sec:LCPRM}. In this case, the decoder queries all elements corresponding to $L_\mathbf{w}(\mathbf{v})\setminus \mathbf{w}$, and employs the extended RS decoding algorithm to decode the symbol. Precisely, assume there exists a codeword generated by $F(\mathbf{X})$. The decoder received the codeword $\mathbf{y}$ with $\delta n$ errors and a point $\mathbf{w}\in \mathcal{N}(\mathbb{P}^{m+1})$, the decoder try to recover $F(\mathbf{w})$ by accessing at most $q$ symbols of $\mathbf{y}$. The algorithm consists of two steps. In the first step, 
\[
L_\mathbf{w}(\mathbf{v}):=\{\mathcal{N}(\mathbf{w}+\lambda\mathbf{v})|\lambda\in \mathbb{F}_q^*\cup \{\infty\}\}.
\]
is constructed by choosing $\mathbf{v}\in \mathcal{N}(\mathbb{P}^{m+1})\setminus \mathbf{w}$ in uniform distribution. Then the codeword symbols indexed by $L_\mathbf{w}(\mathbf{v})$ are queried, and the obtained values are denoted as 
\[
\{e_\lambda=F(\mathcal{N}(\mathbf{w}+\lambda\mathbf{v}) )|\lambda\in \mathbb{F}_q^*\cup \{\infty\}\}.
\]
In the second step, the local decoder try to finds out a univariate polynomial $H(X)$ with $\deg H\leq d$, such that 
\[
H(\lambda)=D_\lambda^d\times  e_\lambda \qquad \lambda\in \mathbb{F}_q^*\cup \{\infty\},
\]
can be satisfied as many as possible, where $D_\lambda=\mathcal{D}(\mathbf{w}+\mathbf{v}\cdot \lambda)$. If $H(X)$ can be determined, the decoder outputs $H(0)$, or else outputs error. By RS decoders, it is known that if the number of unsatisfied equations~(errors) are smaller than $\lfloor (1-\sigma)q/2\rfloor$, the polynomial can be uniquely determined. 

As each query set is individual, the probability lower bound of the successful decoding can be evaluated by Markov inequality. This gives that the probability that $H(X)$ cannot be determined is at most $2\delta/(1-\sigma)$.
\end{proof}
\section{Conclusion}\label{sec:con}
This paper shows that PRM codes form a family of LCCs in the regime of low query complexities. When $q=2$ and $d=1$, PRM codes are Hadamard codes, and the proposed local decoder is the same with the known decoder for Hadamard codes. Further, given a class of GRM codes, we show that there exists a class of PRM codes that is shorter than RM codes with the same query complexities and message lengths.

\bibliographystyle{IEEEtran}
\bibliography{IEEEabrv,refs}

\begin{thebibliography}{10}
\providecommand{\url}[1]{#1}
\csname url@samestyle\endcsname
\providecommand{\newblock}{\relax}
\providecommand{\bibinfo}[2]{#2}
\providecommand{\BIBentrySTDinterwordspacing}{\spaceskip=0pt\relax}
\providecommand{\BIBentryALTinterwordstretchfactor}{4}
\providecommand{\BIBentryALTinterwordspacing}{\spaceskip=\fontdimen2\font plus
\BIBentryALTinterwordstretchfactor\fontdimen3\font minus
  \fontdimen4\font\relax}
\providecommand{\BIBforeignlanguage}[2]{{%
\expandafter\ifx\csname l@#1\endcsname\relax
\typeout{** WARNING: IEEEtran.bst: No hyphenation pattern has been}%
\typeout{** loaded for the language `#1'. Using the pattern for}%
\typeout{** the default language instead.}%
\else
\language=\csname l@#1\endcsname
\fi
#2}}
\providecommand{\BIBdecl}{\relax}
\BIBdecl

\bibitem{Lachaud:1988}
\BIBentryALTinterwordspacing
G.~Lachaud, ``Projective {Reed-Muller} codes,'' in \emph{On Coding Theory and
  Applications}.\hskip 1em plus 0.5em minus 0.4em\relax London, UK, UK:
  Springer-Verlag, 1988, pp. 125--129. [Online]. Available:
  \url{http://dl.acm.org/citation.cfm?id=60380.60393}
\BIBentrySTDinterwordspacing

\bibitem{104317}
A.~B. S{\o}rensen, ``Projective {Reed-Muller} codes,'' \emph{IEEE Transactions
  on Information Theory}, vol.~37, no.~6, pp. 1567--1576, Nov 1991.

\bibitem{Berger2001}
T.~P. Berger and L.~de~Maximy, ``Cyclic projective {Reed-Muller} codes,'' in
  \emph{In Proc. of Applied Algebra, Algrbraic Algorithms and Error Correcting
  Codes}.\hskip 1em plus 0.5em minus 0.4em\relax London, UK, UK: Springer,
  2001, pp. 77--81.

\bibitem{Ding2002}
\BIBentryALTinterwordspacing
P.~Ding and J.~D. Key, ``Subcodes of the projective generalized {Reed-Muller}
  codes spanned by minimum-weight vectors,'' \emph{Des. Codes Cryptography},
  vol.~26, no. 1-3, pp. 197--211, Jun. 2002. [Online]. Available:
  \url{http://dx.doi.org/10.1023/A:1016517611818}
\BIBentrySTDinterwordspacing

\bibitem{995540}
T.~P. Berger, ``Automorphism groups of homogeneous and projective {Reed-Muller}
  codes,'' \emph{IEEE Transactions on Information Theory}, vol.~48, no.~5, pp.
  1035--1045, May 2002.

\bibitem{Ballet2014}
\BIBentryALTinterwordspacing
S.~Ballet and R.~Rolland, ``On low weight codewords of generalized affine and
  projective {Reed--Muller} codes,'' \emph{Designs, Codes and Cryptography},
  vol.~73, no.~2, pp. 271--297, 2014. [Online]. Available:
  \url{http://dx.doi.org/10.1007/s10623-013-9911-7}
\BIBentrySTDinterwordspacing

\bibitem{7572207}
C.~Carvalho and V.~G. Neumann, ``The next-to-minimal weights of binary
  projective {Reed-Muller} codes,'' \emph{IEEE Transactions on Information
  Theory}, vol.~PP, no.~99, pp. 1--1, 2016.

\bibitem{6979865}
N.~Nakashima and H.~Matsui, ``A decoding algorithm for projective {Reed-Muller}
  codes of 2-dimensional projective space with {DFT},'' in \emph{Information
  Theory and its Applications (ISITA), 2014 International Symposium on}, Oct
  2014, pp. 358--362.

\bibitem{NAKASHIMA2016}
------, ``Decoding of projective {Reed-Muller} codes by dividing a projective
  space into affine spaces,'' \emph{IEICE Transactions on Fundamentals of
  Electronics, Communications and Computer Sciences}, vol. E99.A, no.~3, pp.
  733--741, 2016.

\bibitem{TCS-030}
\BIBentryALTinterwordspacing
S.~Yekhanin, ``Locally decodable codes,'' \emph{Foundations and Trends® in
  Theoretical Computer Science}, vol.~6, no.~3, pp. 139--255, 2012. [Online].
  Available: \url{http://dx.doi.org/10.1561/0400000030}
\BIBentrySTDinterwordspacing

\bibitem{1054127}
T.~Kasami, S.~Lin, and W.~Peterson, ``New generalizations of the {Reed-Muller}
  codes--i: Primitive codes,'' \emph{IEEE Transactions on Information Theory},
  vol.~14, no.~2, pp. 189--199, Mar 1968.

\bibitem{6499441}
D.~E. Muller, ``Application of boolean algebra to switching circuit design and
  to error detection,'' \emph{Transactions of the I.R.E. Professional Group on
  Electronic Computers}, vol. EC-3, no.~3, pp. 6--12, Sept 1954.

\bibitem{1057465}
I.~Reed, ``A class of multiple-error-correcting codes and the decoding
  scheme,'' \emph{Transactions of the IRE Professional Group on Information
  Theory}, vol.~4, no.~4, pp. 38--49, September 1954.

\bibitem{Yekhanin:2008}
\BIBentryALTinterwordspacing
S.~Yekhanin, ``Towards 3-query locally decodable codes of subexponential
  length,'' \emph{J. ACM}, vol.~55, no.~1, pp. 1:1--1:16, Feb. 2008. [Online].
  Available: \url{http://doi.acm.org/10.1145/1326554.1326555}
\BIBentrySTDinterwordspacing

\bibitem{Dvir:2010}
\BIBentryALTinterwordspacing
Z.~Dvir, P.~Gopalan, and S.~Yekhanin, ``Matching vector codes,'' in
  \emph{Proceedings of the 2010 IEEE 51st Annual Symposium on Foundations of
  Computer Science}, ser. FOCS '10.\hskip 1em plus 0.5em minus 0.4em\relax
  Washington, DC, USA: IEEE Computer Society, 2010, pp. 705--714. [Online].
  Available: \url{http://dx.doi.org/10.1109/FOCS.2010.73}
\BIBentrySTDinterwordspacing

\bibitem{Guo:2013}
\BIBentryALTinterwordspacing
A.~Guo, S.~Kopparty, and M.~Sudan, ``New affine-invariant codes from lifting,''
  in \emph{Proceedings of the 4th Conference on Innovations in Theoretical
  Computer Science}, ser. ITCS '13.\hskip 1em plus 0.5em minus 0.4em\relax New
  York, NY, USA: ACM, 2013, pp. 529--540. [Online]. Available:
  \url{http://doi.acm.org/10.1145/2422436.2422494}
\BIBentrySTDinterwordspacing

\bibitem{Kopparty:2014}
\BIBentryALTinterwordspacing
S.~Kopparty, S.~Saraf, and S.~Yekhanin, ``High-rate codes with sublinear-time
  decoding,'' \emph{J. ACM}, vol.~61, no.~5, pp. 28:1--28:20, Sep. 2014.
  [Online]. Available: \url{http://doi.acm.org/10.1145/2629416}
\BIBentrySTDinterwordspacing

\bibitem{Kopparty:2016}
\BIBentryALTinterwordspacing
S.~Kopparty, O.~Meir, N.~Ron-Zewi, and S.~Saraf, ``High-rate
  locally-correctable and locally-testable codes with sub-polynomial query
  complexity,'' in \emph{Proceedings of the 48th Annual ACM SIGACT Symposium on
  Theory of Computing}, ser. STOC 2016.\hskip 1em plus 0.5em minus 0.4em\relax
  New York, NY, USA: ACM, 2016, pp. 202--215. [Online]. Available:
  \url{http://doi.acm.org/10.1145/2897518.2897523}
\BIBentrySTDinterwordspacing

\bibitem{Trevisan04}
L.~Trevisan, ``Some applications of coding theory in computational
  complexity,'' \emph{Quaderni di Matematica}, vol.~13, p. 2004, 2004.

\bibitem{Hemenway13}
B.~Hemenway, R.~Ostrovsky, and M.~Wootters, ``Local correctability of expander
  codes,'' in \emph{In Automata, Languages, and Programming}.\hskip 1em plus
  0.5em minus 0.4em\relax Springer, 2013, pp. 540--551.

\bibitem{1960}
I.~S. Reed and G.~Solomon, ``Polynomial codes over certain finite fields,''
  \emph{Journal of the Society for Industrial and Applied Mathematics}, vol.~8,
  no.~2, pp. 300--304, 1960.

\bibitem{WelchBerlekamp1986}
L.~R. Welch and E.~R. Berlekamp, ``{Error correction for algebraic block
  codes},'' US Patent 4 633 470, Dec. 1986.

\bibitem{1054260}
J.~Massey, ``Shift-register synthesis and {BCH} decoding,'' \emph{IEEE
  Transactions on Information Theory}, vol.~15, no.~1, pp. 122--127, Jan 1969.

\bibitem{535795}
J.~O. Jensen, ``On decoding doubly extended {Reed-Solomon} codes,'' in
  \emph{Information Theory, 1995. Proceedings., 1995 IEEE International
  Symposium on}, Sep 1995, pp. 280--.

\bibitem{510064}
L.~L. Joiner and J.~J. Komo, ``Time domain decoding of extended {Reed-Solomon}
  codes,'' in \emph{Southeastcon '96. Bringing Together Education, Science and
  Technology., Proceedings of the IEEE}, Apr 1996, pp. 238--241.

\bibitem{BLAKE19751}
\BIBentryALTinterwordspacing
I.~F. Blake and R.~C. Mullin, ``1 - finite fields and coding theory,'' in
  \emph{The Mathematical Theory of Coding}, I.~F. Blake and R.~C. Mullin,
  Eds.\hskip 1em plus 0.5em minus 0.4em\relax Academic Press, 1975, pp. 1 --
  94. [Online]. Available:
  \url{http://www.sciencedirect.com/science/article/pii/B9780121035501500060}
\BIBentrySTDinterwordspacing

\end{thebibliography}

\end{document}